\documentclass[%
 amsmath,amssymb,
 aps,
]{revtex4-1}

\usepackage{graphicx}
\usepackage{dcolumn}
\usepackage{bm}

\begin{document}

\title{Compact Chiral Boson Fields on the Horizon of BTZ Black Hole}
\author{Jingbo Wang}
\email{ shuijing@mail.bnu.edu.cn}
\affiliation{Institute for Gravitation and Astrophysics, College of Physics and Electronic Engineering, Xinyang Normal University, Xinyang, 464000, P. R. China}
 \date{\today}
\begin{abstract}
In the previous work, it was shown that the degrees of freedom on the horizon of BTZ black hole can be described by two chiral massless scalar fields with opposite chirality. In this paper, we continuous this research. It is found that the scalar field is actually a compact boson field on a circle. The compactness results in the quantization of the black hole radius. Then we quantize the two boson fields and get two abelian Kac-Moody algebras. From the boson field, one can construct the full $W_{1+\infty}$ algebra which was used to classify the BTZ black holes.
\end{abstract}
\maketitle
\section{Introduction}
Black hole physics is a hot spot in theoretical physics since a long time ago. The pioneering works of Bekenstein \cite{bk1}, Hawking \cite{hawk1} and others \cite{bch1} during the seventies of last century have suggested that black holes have temperature and entropy, and satisfy thermodynamics laws. The entropy is given by the famous Bekenstein-Hawking area law. Understanding those properties is a fundamental challenge of quantum gravity, especially the black hole microstates which account for the entropy.

For generic black holes in three dimensional spacetime (the BTZ black hole \cite{btz1}), an explicit proposal for all microstates was given in Ref.\cite{fluff1}. The proposal bases on the so-called ``horizon fluff" \cite{fluff1,fluff2,fluff3}. From a new near horizon boundary condition \cite{bc1,bc2}, a new near horizon symmetry algebra, which is infinite copies of the Heisenberg algebra was obtained. The horizon fluff forms a finite subset of the related ``soft Heisenberg hair". Use this algebra one can generate descendants of physical states which are interpreted as black hole microstates in three dimensional spacetime. The number of those microstates counts for the entropy of the black holes.

In the previous papers \cite{wangbms1,wangbms2}, we use $W_{1+\infty}$ algebra instead of the infinite copies of the Heisenberg algebra to give the BTZ black hole microstates. This $W_{1+\infty}$ was borrowed from the symmetry of the quantum Hall fluid. It was claim that \cite{wangti1,wangti2} the BTZ black hole can be considered as quantum spin Hall state, which is equivalent to bilayer quantum Hall system with opposite $T-$ symmetry. Those incompressible quantum fluids and their excitations can be viewed as $W_{1+\infty}$ edge conformal field theories, thus providing an algebraic classification of quantum Hall universality classes \cite{cap1,cap2}. The $W_{1+\infty}$ was used to classify the BTZ black holes and give the `W-hairs' of black hole \cite{whair1}, which maybe essential to solve the information paradox \cite{info1,info2,info3}. But the origin of this $W_{1+\infty}$ is unclear. In this paper, we show that the $W_{1+\infty}$ algebra can be obtained from the compact chiral boson field on the horizon of BTZ black hole.

It was shown that the degrees of freedom on the BTZ black hole horizon can be described by two chiral massless scalar field theory with opposite chirality \cite{whcft1}. The action for those scalar fields was obtained from the WZW theory restricted on the horizon. In this paper, the scalar field is studied more in-depth. Firstly we find that the scalar field is actually a compact boson field. This gives an important result: the related center-of-mass momentum is quantized. For BTZ black hole, it leads to the quantization of the radius of the black hole. Then we quantized those boson fields to get two abelian Kac-Moody algebras. The Hilbert space forms the representation of those Kac-Moody algebras. The highest-weight states correspond to the BTZ black holes. Finally we construct the $W_{1+\infty}$ algebra from the boson field.

The paper is organized as follows. In section II, it is shown that the scalar field is actually a compact boson field on the horizon. In section III, we quantize those boson fields and get two abelian Kac-Moody algebras. The highest-weight states for representation are also given. From the boson field, one can construct the $W_{1+\infty}$ algebra. Section IV is the conclusion.
\section{The Compact Chiral Boson on the Horizon}
As first shown in Ref.\cite{at1}, $(2+1)-$dimensional general relativity can be written as a Chern-Simons theory. For the case of negative cosmological constant $\Lambda=-1/L^2$, one can define two SO$(2,1)$ connection 1-forms
\begin{equation}\label{1}
    A^{(\pm)a}=\omega^a\pm \frac{1}{L} e^a,
\end{equation}where $e^a$ and $\omega^a$ are the orthonormal co-triads and spin connection 1-forms, respectively, and $a=0,1,2$ is gauge group index. Then, up to a boundary term, the first order action of gravity can be rewritten as
\begin{equation}\label{2}\begin{split}
    I_{GR}[e,\omega]=\frac{1}{8\pi G}\int e^a \wedge ({\rm d}\omega_a+\frac{1}{2}\epsilon_{abc}\omega^b \wedge \omega^c)-\frac{1}{6L^2}\epsilon_{abc}e^a\wedge e^b \wedge e^c\\=I_{CS}[A^{(+)}]-I_{CS}[A^{(-)}],
\end{split}\end{equation}
where $A^{(\pm)}=A^{(\pm)a}T_a$ are SO$(2,1)$ gauge potential, $T_a$ are generators of SO$(2,1)$ group, and the Chern-Simons action is
\begin{equation}\label{3}
    I_{CS}[A]=\frac{k}{4\pi}\int {\rm Tr}\{A\wedge {\rm d}A+\frac{2}{3}A\wedge A \wedge A\},
\end{equation}
with
\begin{equation}\label{4}
    k=\frac{L}{4G}.
\end{equation}
Similarly, the CS equation
\begin{equation}\label{5}
    F^{(\pm)}={\rm d}A^{(\pm)}+A^{(\pm)} \wedge A^{(\pm)}=0
\end{equation}
is equivalent to the requirement that the connection is torsion-free and the metric has a constant negative Riemann curvature.
The equation implies that the potential $A$ can be locally written as
\begin{equation}\label{5a}
    A=g^{-1}{\rm d}g,
\end{equation}
where $g$ is an SO$(2,1)$ group element. Using the Gauss decomposition, it can be written as
\begin{equation}\label{5b}
    g=\left(
                  \begin{array}{cc}
                    1 & \frac{1}{\sqrt{2}}x \\
                    0 & 1 \\
                  \end{array}
                \right)
                \left(
                         \begin{array}{cc}
                           e^{\Psi/2} & 0 \\
                           0 & e^{-\Psi/2} \\
                         \end{array}
                       \right)
                       \left(
                         \begin{array}{cc}
                           1 & 0 \\
                           -\frac{1}{\sqrt{2}}y & 1 \\
                         \end{array}
                       \right).
\end{equation}

To study the physics at a horizon $\Delta$, it is more suitable to use advanced Eddington coordinate $(v,r,\varphi)$. The metric of BTZ black hole can be written as
\begin{equation}\label{6}
    ds^2=-N^2 dv^2+2 dv dr+r^2 (d\varphi+N^\varphi dv)^2.
\end{equation}

In the previous work \cite{whcft1}, we get the result that the BTZ horizon can support two chiral massless scalar fields $(\Psi_1(\tilde{u}),\Psi_2(u))$ with opposite chirality, where $ u=\varphi-\frac{v}{L},\tilde{u}=\varphi+\frac{v}{L}$.The $\Psi$ field is part of SO$(2,1)$ group element $g$ (\ref{5b}).

Since those two fields have similar property, let's first consider the $\Psi_1(\tilde{u})$. In canonical formula, the connection $A_i=g^{-1}\partial_i g$ will solve the constraint $ F_{ij}=0$. Since the horizon $\Delta$ has non-trivial topology, the $g$ is non-periodic, and can be represented by \cite{holo1,rs1}
\begin{equation}\label{7}
  g(r,\varphi,v)=\exp[\varphi h_1] q(r,\varphi,v),
\end{equation}
where $h_1$ is non-trivial flat connection and can be chosen as $h_r=0, h_\varphi=h_1(v)$, and $q(r,\varphi+2\pi,v)=q(r,\varphi,v)$. We choose $h_1(v)$ to be diagonal, i.e. $h_1(v)=h(v) \sigma_z$, and the relation (\ref{5b}) between $g$ and $\Psi_1$ gives
\begin{equation}\label{8}
 \Psi_1(r,\varphi+2\pi,v)=\Psi_1(r,\varphi,v)-2 \pi h(v),
\end{equation}
thus, $\Psi_1$ is actually a compactified boson field.

The $h(v)$ can be calculated through the holonomy of the connection $A$ at fixed time,
\begin{equation}\label{9}
  W=P \exp(\oint A),
\end{equation}
which is related to $g$ by
\begin{equation}\label{10}
  W=g^{-1}(v,r,\varphi=0)g(v,r,\varphi=2\pi)=q^{-1}(v,r,\varphi=0)\exp(2\pi h(v))q(v,r,\varphi=0).
\end{equation}
A straightforward calculation gives
\begin{equation}\label{11}
 h(v)=-\sqrt{2 G (M-J/L)}=-\frac{r_+-r_-}{2L}=-\pi T_L.
\end{equation}
where
\begin{equation}\label{11a}
  T_{R/L}=\frac{r_+\pm r_-}{2\pi L}
\end{equation}
are dimensional right- and left-temperature.

The similar method can be applied to the $\Psi_2(u)$ and gives
\begin{equation}\label{12}
 \Psi_2(r,\varphi+2\pi,v)=\Psi_2(r,\varphi,v)-2 \pi \tilde{h}(v),
\end{equation}
where
\begin{equation}\label{13}
 \tilde{h}(v)=\sqrt{2 G (M+J/L)}=\frac{r_++r_-}{2L}=\pi T_R.
\end{equation}

The entropy can be written as
\begin{equation}\label{14}
  S=\frac{2 \pi r_+}{4 G}=2\pi k ( \tilde{h}(v)-h(v))=2\pi^2 k ( T_R+T_L).
\end{equation}
\section{Canonical quantization of compact chiral boson field}
The action for the left-moving scalar field on the horizon is \cite{whcft1}
\begin{equation}\label{15}\begin{split}
    I&=\frac{k}{4\pi}\int_{\Delta}du d\tilde{u}\frac{1}{2}\partial_u \Psi_1 \partial_{\tilde{u}} \Psi_1\\
    &=\frac{k}{4\pi L}\int_{\Delta} dv d\varphi[L^2 (\partial_v \Psi_1)^2- (\partial_{\varphi} \Psi_1)^2],
   \end{split}\end{equation}
with $\Psi_1$ depending only on $\tilde{u}=\varphi+\frac{v}{L}$. Following the method in Ref.\cite{chiral1}, we quantize this compact chiral boson field.

The Lagrangian density is given by
\begin{equation}\label{16}\begin{split}
    \mathcal{L}=\frac{k L}{4\pi}[(\partial_v \Psi_1)^2- \frac{1}{L^2}(\partial_{\varphi} \Psi_1)^2],
\end{split}\end{equation}
Since $\Psi_1$ is chiral, it satisfy the constraint
\begin{equation}\label{17}
  (\partial_v-\frac{1}{L}\partial_\varphi)\Psi_1=0.
\end{equation}
The general solution to this equation and the boundary condition (\ref{8}) is
\begin{equation}\label{18}
  \Psi_1(\varphi+\frac{v}{L})= \Psi_{10}-\alpha_0(\varphi+\frac{v}{L})+i\sum_{n\neq 0}\frac{\alpha_n}{n}e^{in(\varphi+\frac{v}{L})},\quad \alpha_n^*=\alpha_{-n},
\end{equation}
where $\alpha_0=h(v)=-\pi T_L$.

The conjugate momentum is
\begin{equation}\label{19}
  \pi_1(\varphi+\frac{v}{L})=\frac{\partial \mathcal{L}}{\partial (\partial_v \Psi_1)}= -\frac{k}{2\pi}(\alpha_0+\sum_{n\neq 0}\alpha_n e^{in(\varphi+\frac{v}{L})}).
\end{equation}
The canonical commutation relation $[\Psi_1(\varphi,v),\pi_1(\varphi',v)=i \delta(\varphi-\varphi')$ gives
\begin{equation}\label{20}
  [\Psi_{10},\alpha_0]=-\frac{i}{2k}, \quad [\alpha_n, \alpha_m]=-\frac{n}{2k} \delta_{n+m,0}.
\end{equation}
 The second relation is just anti-chiral part of the abelian Kac-Moody algebra \cite{wangbms2}. Define the new operators
\begin{equation}\label{20a}
\alpha^-_n \equiv \alpha_{-n},
\end{equation}
which indeed satisfy the standard abelian Kac-Moody algebra
\begin{equation}\label{20b}
 [\alpha^-_n, \alpha^-_m]=\frac{n}{2k} \delta_{n+m,0}, \quad [\Psi_{10},\alpha^-_0]=-\frac{i}{2k}.
\end{equation}

 The consistency quantization forces $ k=\frac{L}{4G}$ to be an integer \cite{chiral1}. Since $\Psi_1$ is a compact boson, and $2 k \alpha_0$ is the canonical momentum conjugate to the angular variable $\Psi_{10}$, the momentum has to satisfy the quantization condition
 \begin{equation}\label{21}
   \alpha_0^-=-\frac{n^-}{2k}, \quad n^- \in N.
 \end{equation}
So the temperature is also quantized
\begin{equation}\label{22}
   T_L=\frac{n^-}{2 \pi k}=n^-\frac{2 G}{\pi L}, \quad n^- \in N.
 \end{equation}

The same process can be applied to $\Psi_2$. We write
\begin{equation}\label{23}
  \Psi_2(\varphi-\frac{v}{L})= \Psi_{20}-\alpha^+_0(\varphi-\frac{v}{L})+i\sum_{n\neq 0}\frac{\alpha^+_n}{n}e^{in(\varphi-\frac{v}{L})},\quad (\alpha^+_n)^*=\alpha^+_{-n},
\end{equation}
where $\alpha^+_0=\tilde{h}(v)=\pi T_R$.

The commutation relations read
\begin{equation}\label{24}
  [\Psi_{20},\alpha^+_0]=\frac{i}{2k}, \quad [\alpha^+_n, \alpha^+_m]=\frac{n}{2k} \delta_{n+m,0}.
\end{equation}
The second relation is just the chiral part of Kac-Moody algebra.

We also have the quantization condition
\begin{equation}\label{25}
   \alpha^+_0=\frac{n^+}{2k},\quad  T_R=\frac{n^+}{2 \pi k}=n^+\frac{2 G}{\pi L}, \quad n^+ \in N.
 \end{equation}
They are equivalent to the conditions
\begin{equation}\label{25a}
  r_+=2(n^++n^-)L_{PL}, \quad r_-=2(n^+-n^-)L_{PL},
\end{equation}
where $L_{PL}=G$ is the Planck length. It means that the radius of black hole is quantized.

The entropy have a simple form,
\begin{equation}\label{26}
  S=2\pi k ( \tilde{h}(v)-h(v))=\pi (n^++n^-).
\end{equation}

The Virasoro operators can be defined through Sugawara construction
\begin{equation}\label{27}\begin{split}
  L^\pm_0=k (\alpha^\pm_0)^2+2 k \sum_{n=1}^{\infty} :\alpha^\pm_{-n}\alpha^\pm_n:,\\
  L^\pm_n=k \sum_{l=-\infty}^{+\infty} :\alpha^\pm_{n-l}\alpha^\pm_l:,
\end{split}\end{equation}
where $::$ is normal ordering. Those operators satisfy the Virasoro algebra with central charge $c=1$,
\begin{equation}\label{27a}\begin{split}
    [L^\pm_n, \alpha^\pm_m]&=-m \alpha^\pm_{n+m},\\
  [L^\pm_n, L^\pm_m]&=(n-m)L^\pm_{n+m}+\frac{1}{12} n(n^2-1)\delta_{n+m,0}.\\
\end{split}\end{equation}

Each highest-weight state can be parametrized by the eigenvalues of $\alpha^\pm_0$, that is the $|n^+,n^->$. The relation with the absolute vacuum $|0,0>$ is given by the vertex operator \cite{cft1}
\begin{equation}\label{28}
  |n^+,n^->=:e^{\frac{i}{2k} (n^+ \Psi_2-n^- \Psi_1)}: |0,0>
\end{equation}
with the properties
\begin{equation}\label{29}
  \alpha^\pm_n |n^+,n^->=0 \quad (n>0),\quad \alpha^\pm_0 |n^+,n^->=\pm \frac{n^\pm}{2k} |n^+,n^->.
\end{equation}

Define Hamiltonian and angular momentum operator respectly,
\begin{equation}\label{32}
  \hat{H}=L^+_0+L_0^-,\quad \hat{J}=L^+_0-L_0^-,
\end{equation}
then it is easy to find that
\begin{equation}\label{33}
  <n^+,n^-|\hat{J}|n^+,n^->=\frac{G}{L}((n^+)^2-(n^-)^2)=J,\quad <n^+,n^-|\hat{H}|n^+,n^->=\frac{G}{L}((n^+)^2+(n^-)^2)=M L,
\end{equation}
where $J,M$ are just the angular momentum and the mass of the BTZ black hole. So the highest-weight state corresponds to BTZ black hole.

Finally, from the chiral boson field, one can get the full $W_{1+\infty}$ algebra \cite{cap1,cap2,az1}. First one use the equivalence between chiral boson field $\Psi$ and Weyl fermion field $\psi$ in $1+1$ dimensional spacetime \cite{fj1}. The Hamiltonian of fermion field $\psi$ is
\begin{equation}\label{34}
  H=\int dx \psi^+ (i\partial_x) \psi +h.c.,
\end{equation}
and the generators of the $W_{1+\infty}$ algebra $V^j_n$ can be represented by
\begin{equation}\label{35}
  V^j_n=\int_0^{2 \pi} d\theta \psi^+(\theta) \ddag e^{-i n \theta}(-i\partial_\theta)^j\ddag \psi(\theta),
\end{equation}
where $\ddag \ddag$ denotes an ordering such that $(V^j_n)^+=V^j_{-n}$. Those generators satisfy the $W_{1+\infty}$ algebra with central charge $c=1$.

Since we have two chiral boson fields with opposite chirality, the final symmetry algebra should be $W_{1+\infty} \otimes \bar{W}_{1+\infty}$. This algebra was used to gives the W-hairs of BTZ black holes \cite{wangbms2}.
\section{Conclusion}
In this paper, we study the compact chiral boson fields on the horizon of BTZ black hole. The quantum theory give two copies of abelian Kac-Moody algebras. The highest-weight states are identified with BTZ black holes. The compactness results in the quantization of the radius of black hole. From the chiral boson field one can construct $W_{1+\infty}$ algebra which can be used to classify the BTZ black holes.

It is helpful to compare with the ``horizon fluff" approach \cite{fluff1,fluff2,fluff3}. In their approach, starting from the near horizon Kac-Moody algebra, one can get a free boson theory on the horizon. But they omit the first commutation relation of (\ref{20}) (the zero-mode) which was already noted in Ref.\cite{fluff2}. We start from the action of the compact chiral boson field and quantize this field to get the Kac-Moody algebra plus the zero-mode commutation relation. The compactness results in the quantization of the radius of the black holes. Also note that the $\Psi$ field in our approach is closely related with the $\Phi$ field in Ref.\cite{fluff3} in horizon fluff approach.

As a final remark, the algebras (\ref{20a},\ref{24},\ref{27a}) are closely related with the enhanced asymptotic symmetry algebra of $AdS_3$ \cite{ads3}. After the Ba$\tilde{n}$ados map \cite{c1}
\begin{equation}\label{36}
 \tilde{L}^\pm_n=\frac{1}{c}(L^\pm_{cn}-\frac{1}{24}\delta_{n,0}),\quad n \in Z, c\in N,
\end{equation}
and the identities
\begin{equation}\label{37}
  P^\pm_n=\frac{\sqrt{2} k}{\sqrt{c}}\alpha^\pm_{nc}, \quad Q=\frac{\sqrt{c} k}{\sqrt{2}}(\Psi_{20}-\Psi_{10}),
\end{equation}
it is straightforward to show that they just give the enhanced asymptotic symmetry algebra (7.9) of $AdS_3$ in Ref.\cite{ads3}.

\acknowledgments
 This work is supported by the NSFC (Grant No.11647064) and Nanhu Scholars Program for Young Scholars of XYNU.
 \bibliography{it4}
\end{document}